\def   \newline {\hfil\break}

\def\Epk{E_{\rm pk}}
\def\Ep{E_{\rm p}}

\def\E0{E_{\rm 0}}
\def\P0{\Phi_{\rm 0}}
\def\Ep0{E_{\rm p0}}

\def\N0{N_{\rm 0}}
\def\t0{\t_{\rm 0}}

\def\Gt{{\bf G}(t)}

\newcommand{\ltsima} {$\; \buildrel < \over \sim \;$}
\newcommand{\gtsima} {$\; \buildrel > \over \sim \;$}
\newcommand{\lta} {\lower.5ex\hbox{\ltsima}}
\newcommand{\gta} {\lower.5ex\hbox{\gtsima}}

\documentstyle[11pt,paspconf,epsfig]{article}

\begin{document}

\title{Spectral Aspects of the Evolution of Gamma-Ray Bursts}

\author{Felix Ryde}
\affil{Stockholm Observatory, SE-133 36 Saltsj\"obaden, Sweden}

\begin{abstract}
A review on the spectral and temporal properties of 
gamma-ray bursts is given.
Special attention is paid to the spectral evolution of
their continuum emission and
its connection to the time evolution of the intensity.
Efforts on systematizing these observations as well as
the effects due the limitations of the
current detectors on the observed sample are discussed.
Finally, physical models that aim at explaining the observations,
are addressed.
\end{abstract}


\section{Introduction}

The discovery of gamma-ray bursts (GRBs) at the end of the 1960's
(Klebesadel, Strong, \& Olson 1973) revealed a phenomenon which has
been unwilling to allow us to gain a clear insight into its origin.
Until the recent attention given to the GRB afterglow emission
(see, e.g., Metzger et~al. 1997) the prompt non-thermal flash of 
gamma-rays  has been the main source of 
information. The  data collected by the {\it Compton Gamma-Ray 
Observatory, CGRO} (Fishman et~al. 1989) and the {\it BeppoSAX}
(Boella et~al. 1997) satellites, have given an unprecedented wealth 
of data. This has led, after an initial phase of confusion, 
to a more detailed knowledge, for instance,  of the spectral shape
 and its evolution in time.
The observed gamma-ray light curves exhibit a large diversity 
in duration, strength, and morphology. Some are very complex, having 
stochastic spiky structures,
while others are smooth and have only  a few, well-shaped
pulses. The duration of the gamma-ray emission ranges 
from as short as a few milliseconds up to several hundreds of seconds.
The spectra, even though not as
diverse in their characteristics as the light curves, have not  
given any clear signature of the underlying physical
emission process(es).  The spectra have a non-thermal appearance and 
can evolve considerably during the burst.
Undoubtably, the key to the understanding of the
phenomenon lies in this spectral behavior.

Is the large diversity mainly due to
varying physical properties of the source or is it due to other
effects such as different appearances to the observer?  
Are there any characteristics that can 
correctly describe the GRB temporal behavior
and do these have typical values for all GRBs. In other words, 
how broad are their true, intrinsic distributions?
Much study has been devoted to the search for
empirical relations and correlations 
between observable quantities, and 
to systematize the diverse appearances of
the data.
Correlations for both large ensembles of GRBs and within 
individual bursts have been studied.
This is a natural step in astronomy and 
can be compared to the early advances
in our understanding of stellar evolution using 
the optical color-color diagrams, and, for instance,
using the Hertzsprung-Russell diagram of globular clusters
to determine their ages. 
The behavior of low-mass X-ray binaries 
is studied in X-ray
color-color diagrams leading to the classification 
of the sources into two populations based on their behavior
in  the diagrams: Z sources and atoll sources.
In this review,
the study of the temporal-spectral behavior within individual bursts 
will be addressed. It will mainly concern the efforts to
understand the continuum spectral shape and its evolution in time.
These results should trigger and guide theoretical work and lead to
physical models capable of reproducing the observed features.

In \S 2,  the main features of the light curves and spectra as observed
to date, are summarized followed by a description of the spectral
evolution in \S 3. This is succeeded by a discussion on how the 
spectra and the intensity evolve relative to each other in \S 4.
\S 5 is devoted to a discussion  on different aspects of the 
observations which could affect the
results. Finally, an overview discussion on the constraints put on
the physical models describing the data is given in \S 6.

\section{Burst Properties}

In the following, the photon flux will be denoted by 
 $N(t)$ (photons cm$^{-2}$ s$^{-1}$) 
and its spectrum by $N_{\rm E}(E, t)$ (photons cm$^{-2}$ s$^{-1}$ keV$^{-1}$) 
and correspondingly,  the energy flux  by $F(t)$ (keV cm$^{-2}$
s$^{-1}$). 
Intensity will denote a general flux entity and not necessarily
the intensity-entity usually used in astronomy.
In GRB astronomy the sources are not resolved, thus 
making it less meaningful.
The terms light curve and time history will be used 
synonymously with the time evolution of the intensity.
The hardness of the spectrum will refer to the overall spectral property 
of the burst, mainly the peak energy of the power output.
A power-law spectrum is steep if it is dominated by soft photons and
flatter as the fraction of hard photons increases.

\subsection{Gamma-Ray Burst Light Curves}

\begin{figure}
\centerline{\epsfig{file=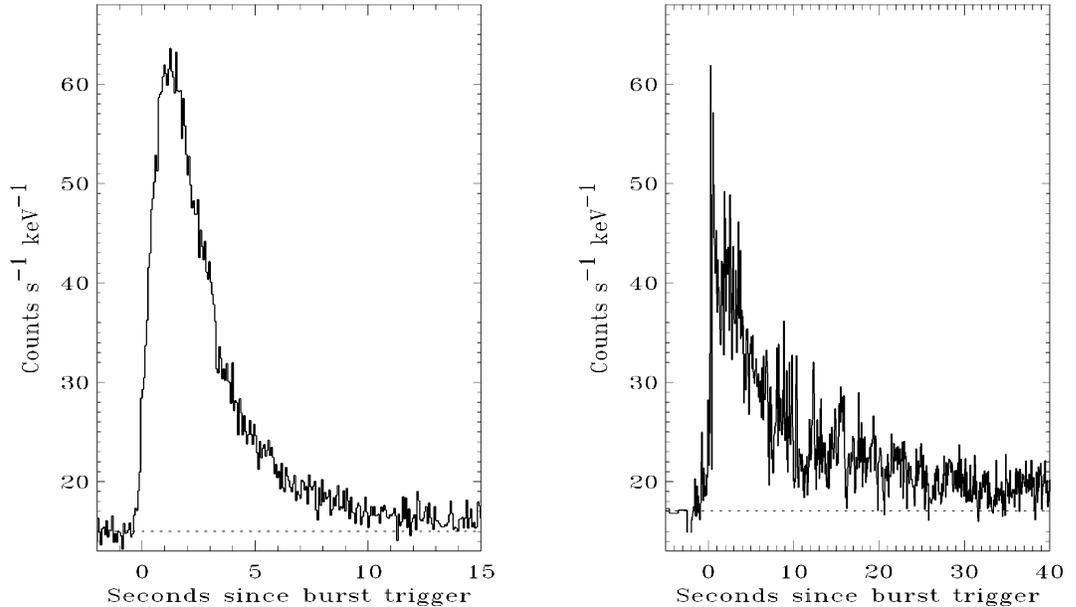,width=14cm,height=8cm}}
\caption{Light curves of GRB 920830 (BATSE trigger \#1883)
 and GRB 940526 (\#2993),
having similar overall envelopes of emission. The dashed line
indicates the background level.}
\label{fig1}
\end{figure}

A remarkable feature of the observed properties
of GRBs is the large diversity of the light curves, both
morphologically and in strength and duration\footnote{This has been 
summarized as: `Have you seen one, you have seen one'}.
Several examples of light curves, observed by the Burst and Transient 
Source Experiment (BATSE) on the {\it CGRO} will be presented 
in this paper (Figure~\ref{fig1}, Figure~\ref{fig43}, and
Figure~\ref{fig10}a; see, e.g., the current
BATSE catalog\footnote{The BATSE GRB catalog is available online at: 
http://www.batse.msfc.nasa.gov/ data/grb/catalog/.}). 
Different approaches to the understanding of the light curve morphology
have been pursued. 
It is generally believed that the
fundamental constituent of a GRB light curve is a time structure
having 
a sharp rise and a slower decay, with the decay rate decreasing
smoothly (e.g., Fishman et~al. 1994; Norris et~al.  1996; Stern \& Svensson
1996).
This shape is denoted by the acronym FRED, fast-rise and
exponential-decay, even though the decay is not necessarily
exponential. A burst can consist of only
a few such pulses, clearly separable, producing a simple and smooth
light curve, as in the left-hand panels in Figure~\ref{fig1} 
and Figure~\ref{fig10}a.  More complex light curves,
such as in Figure~\ref{fig43} are superpositions of many such
fundamental pulses. Mixtures of the two types are also common. 
Such interpretations have been shown to be able to
explain and partly reproduce many observed light curve morphologies.
To reveal the underlying process of GRBs, the fundamental pulses are of extra
interest as they will show the clearest signature of the physics.  
To model pulses, often a stretched exponential is used:
$N(t) \propto $ exp $(-(\vert t-t_{\rm max}\vert /\sigma_{\rm
r,d})^{\nu})$,
where $t_{\rm max}$ is the time of the pulse's maximum intensity,
$\sigma_{\rm r,d}$ are the time constants for the rise and the decay,
and $\nu$ is the peakedness parameter.
Such a function gives a flexibility to describe most pulses,
and to give characteristics of the pulses for statistical analysis.
Norris et~al. (1996) studied a sample of 
bursts observed by the BATSE Large Area Detectors (LADs)
and stored in the four energy channel data type\footnote{For the
different data types from  BATSE, see, e.g., Fishman et~al. (1989)}.
They modeled the light curves in detector counts in the four channels
separately  and
found that the decay generally lay between a pure exponential
($\nu = 1$) and a Gaussian ($ \nu = 2$). Lee et~al. (1998) studied 
approximately 2500 individual channel pulse structures in the high time 
resolution BATSE TTS data, using this general stretched exponential
function and confirmed the general behavior that pulses tend to
have shorter rise times than decay times.
Norris et~al. (1996) also used the stretched
exponential to create an algorithm to separate
overlapping pulses based on $\chi ^2$ fitting.
Another pulse-identification algorithm was
introduced by Li \& Fenimore (1996) and similarly by
Pendleton et~al. (1997), who identified pulses  
based on the depth of the minima between peaks, which has the
advantage that it does not depend on any particular peak shape.

The large amplitude variations observed within a 
burst is demonstrated by Stern (1999) who shows a few examples of 
GRB pulses with near-exponential tails that are traceable over
almost 4 orders of magnitude in intensity. 
Schaefer \& Dyson (1996)  studied the decay phase of 10 smooth FRED
pulses in the four separate energy channels and found that most of them 
are not exponentials, although a few
cases come close. A power-law fit passes most of their statistical 
tests.

Most  studies  use the LAD 4  spectral  channel  data,  with  the  channels
covering  approximately 20-50 keV, 50-100 keV, 100-300 keV, and 300 keV - 2
MeV.  Often  the  studies  make use of  individual  channels  or the sum of
channels  2 and 3  (50-300  keV) in count  space,  without  using  detailed
knowledge  of the  spectral  behavior.  Frequently,  the  count  rates  are
normalized in the four  channels.  It is, however, of interest to study the
intensity  curve in {\it photon flux} for the 
maximal  available  energy  band-width
instead of detector counts to get physical values on the fitted parameters.
This is done by correctly considering the effects of the detector response.
The  spectra  must then be  deconvolved  for every time bin, for  instance,
using direct inversion  techniques, which can be model independent (Loredo
\& Epstein 1989).  Alternatively,  forward-folding  techniques can be used,
fitting an empirical spectral model to the data by minimizing the $\chi ^2$
between the model count spectrum and the observed count  spectrum.  Ryde \&
Svensson  (1999b) used the LAD 128 spectral  channel data, between  25-1900
keV, to study GRB  pulses in {\it photon  flux}.  They  identify  a 
subgroup  of
pulses  for which the early  intensity  decay  follows a  power-law,  $N(t)
\propto  (1+t/\tau)  ^{-1}$, where the time coordinate,  $t$, is taken from
the  maximum  of the light  curve and  $\tau$  is the time  constant.  This
behavior changes  eventually into a faster decay such as an exponential.  A
detailed discussion on this issue is given in $\S 6.4$.

The light curve has also been described as
having an overall envelope with a FRED-like shape
(see, e.g., Fenimore et~al. 1996).
The actual light curve can either 
follow the envelope closely or have more or less strong deviations, giving
rise to a stochastic, spiky appearance, see Figure~\ref{fig1}. 
In that model, even a simple pulse is thus not caused by a single event 
but is the result of several.

There has also been a proposal that the light curve can be decomposed into 
two uncorrelated radiation components, which dominate different
parts of the spectrum and behave differently 
(Chernenko \& Mitrofanov 1995).
The corresponding spectral behavior has also been studied by 
Chernenko et~al. (1998) who model the spectrum with 9 parameters,
describing the two emission components. The authors studied
approximately 10 strong BATSE bursts, which could all be
explained by the model.

In several works, the averaged behavior over the entire burst has
been studied. By aligning the light curves (summed over all four
LAD channels) to the time of the peak of the
event, the averaged, {\it peak-aligned} profile is obtained. This has been
done by, for
instance, Mitrofanov et~al. (1994, 1996), Norris et~al. (1994), Stern
(1996) and Stern et al. (1997, 1999). 
Stern (1996) showed that the averaged, peak-aligned profile has 
an overall stretched exponential form  with the index $\nu = 1/3$.
Another alternative average is the averaged, {\it duration-aligned}
profile, which is obtained by aligning the time structure 
by setting the durations to a standard duration. The decay is then 
found to fit a linear function in time, but not to exponentials and 
power-laws  (Fenimore 1999).

A general remark on the time histories is that there are no typical 
starting points of the 
emission that, for instance, could be associated with the primary event.

\subsection{Gamma-Ray Burst Spectra}

\begin{figure}
\centerline{\epsfig{file=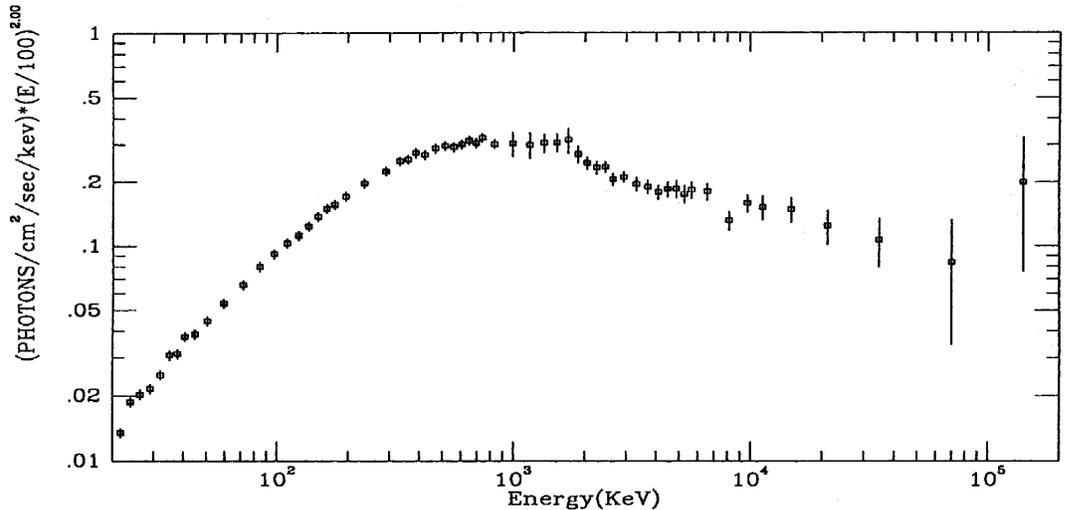,width=14cm,height=7cm}}
\caption{Composite spectrum of GRB 910503 (\# 143) using all
the capability of the {\it CGRO}'s four experiments. From 
Schaefer et~al. (1994).}
\label{fig2}
\end{figure}

\begin{figure}
\centerline{\epsfig{file=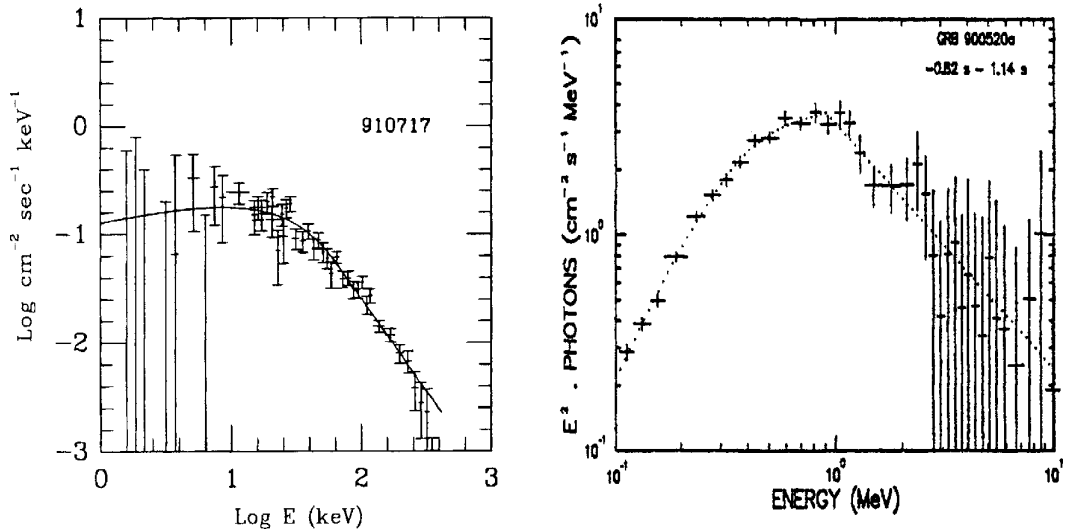,width=14cm,height=7cm}}
\caption{Left panel: Photon spectrum, $N_{\rm E}(E)$, of GRB 910717
observed by {\it Ginga}. From Strohmayer et~al. (1998). 
Right panel: $E ^2 N_{\rm E}(E)$
spectrum of GRB 900520a observed by PHEBUS/{\it Granat}. 
From Barat et~al. (1998).}
\label{fig25}
\end{figure}

\subsubsection{Pre-BATSE Results.}
Important results concerning the GRB spectrum were obtained with a
number of experiments prior to the {\it CGRO}, such as {\it IMP 6} 
(Cline et~al. 1973) and {\it IMP 7} (Cline \& Desai 1975), 
SIGNE/{\it Venera} (Chambon et~al.  1979), 
KONUS/{\it Venera} (Mazets et al. 1982), the GRS/{\it Solar Maximum 
Mission} (Matz et~al.  1985), and {\it Ginga} (Yoshida et~al. 1989). 
The results of these experiments indicated that the
spectral continuum, in the keV to MeV range, 
consisted of three major components: 
(i) a low-energy component resembling the thermal bremsstrahlung 
spectrum (with a Gaunt factor of 1) of an optically-thin hot 
plasma (e.g., Rybicki \& Lightman 1979):
 $N_{\rm E}(E) = E^{-1}$ exp$(-E/E_{\rm 0})$, where $E$ is the 
spectral energy, and $E_{\rm 0}$ is the $e$-folding energy.
(ii) a steep high-energy power-law with no obvious cut-off, $N_{\rm E}(E) 
\propto E^{-2.5}$. Matz et~al. (1985) showed that 60 $\%$ of
their sample had emission above 1 MeV. 
(iii) an X-ray component ($< 10$ keV) which resides 
1-2 $\%$ of the total power.
There were also several reports of emission and absorption features.
Mazets et~al. (1981) presented observations of features
from the KONUS/{\it Venera} 
and Hueter et~al. (1987) from {\it HEAO-1}. Additional reports on 
spectral features  were given by Muakami et~al. (1988) from the 
{\it Ginga} observations.

\subsubsection{BATSE Results.}
The BATSE detectors ($\sim 20-1900$ keV) have given a large 
data base, which refined
these results. Palmer et~al. (1994) studied 192 bursts with the
spectroscopy detectors (SDs). The SDs have the ability to see spectral
lines with the
characteristics reported by Ginga. No convincing line features were found in
the BATSE data,
ruling out the previous results. This result is confirmed by the high 
energy resolution TGRS (transient gamma-ray spectrometer) on the {\it WIND}
spacecraft (Palmer et~al. 1996; Seifert et~al. 1997).
It now seems likely that if line features exist they are very rare.
Briggs et~al. (1998) and Golenetskii et~al. (1998)
report on a few possible candidates from BATSE and KONUS/{\it WIND}, 
respectively.
Pendleton et~al. (1994) found that broad cusps in the energy range
40-100 keV, could be explained by a superposition of hard and soft
spectral sub-components. This was also suggested within the picture
presented in Ryde \& Svensson (1999a). A detailed discussion on
the methodology of identifying line features in gamma-ray spectra
and a review of the field is given in Briggs (1999). 

Schaefer et~al. (1992) and Band et~al. (1993) studied the
continuum spectral shape of the BATSE bursts. The latter study 
comprised a sample of 54 BATSE bursts and  
successfully fitted most spectra with an empirical model
similar to the optically-thin bremsstrahlung spectrum previously used;
a low-energy power-law exponentially joined together with a 
high-energy power-law. The success of this model, fitting both the 
time-integrated spectra and the time-resolved spectra, has led to a wide
spread use and it is often referred to as the `GRB-function' or the 
`Band-model':
\begin{equation}
 N_{\rm E}(E) = \left\{ \begin{array}{ll}
            A \left(\frac{E}{100 \, {\rm keV}}\right)^{\alpha}
e^{-E/E_{\rm 0}} & \mbox{if $ E \leq (\alpha-\beta) E_{\rm 0}$};\\
            A' \left(\frac{E}{100 \, {\rm keV}}\right)^{\beta} & \mbox{if
$E > (\alpha-\beta) E_{\rm 0}$} \end{array} \right. 
\end{equation}
\noindent
where $E_{\rm 0}$ is the $e$-folding energy (in units of keV), and 
\begin{equation}
A'=A \left[ \frac{(\alpha-\beta)E_{\rm 0}}{100 \, {\rm keV}} 
\right]^{\alpha-\beta}
e^{-(\alpha-\beta)},
\end{equation}
\noindent
with  $N_{\rm E}(E)$ being a continuous and a continuously
differentiable function.
Often the energy at which the power 
is maximal, $E_{\rm pk}= (2+\alpha) E_{\rm 0}$ (the peak in the logarithmic 
$E^2 N_{\rm E}$-spectrum), is used as the measure of
spectral hardness, and not $E_{\rm 0}$.
A power peak exists only in
the case of $\beta < -2$.
The Band et~al. (1993) study did not identify any universal values for 
the GRB-function parameters, which were found to have a large
diversity. The peak energy  lies mainly in the interval 100 keV to 1 MeV, 
clustering around 100-200 keV.
The study also confirmed the existence of a hard tail. 
In a recent study by Preece et ~al. (1999), the mean of the distribution of 
the peak energies was
determined to be $E_{\rm pk} = 250 ^{+433} _{-143}$.

The GRB-function index, $\alpha$, shows
a broad peak between $-1.25$ and $-0.25$, 
instead of the universal value earlier claimed of $-1$ (Band
et~al. 1993), while the high-energy power-law index, $\beta$  
clusters fairly narrowly around $-2.12~\pm~0.30$, even though there
exist super-soft bursts with $\beta < -3$ (Preece et~al. 1998a).
Schaefer \& Walker (1999) noted, for instance, that the spectrum 
of GRB 920229 has an extremely sharp high-energy cut-off.
Some studies  have tried to identify statistically
averaged shapes by
various methods of averaging the spectra. Fenimore (1999) 
studied the average spectrum from the duration-aligned light curves
of a sample of GRBs and found  that $\langle\alpha\rangle=-1.03$ and
 $\langle\beta\rangle=-3.31$. The peak energy lied at $390$ keV.

\subsubsection{Outside the BATSE Spectral Range.}
How far do  the power-laws persist towards lower and higher energies? 
Occasionally the GRB lies in the field of view of the other {\it CGRO} 
instruments (COMPTEL, EGRET, OSSE) and a broader 
spectrum can be
studied. An example is given in Figure~\ref{fig2} where a 
composite spectrum of GRB 910503, using all
the capability of the {\it CGRO}'s four experiments, is given.
A similar study was done by Schaefer et al. (1998) who 
studied GRB 910503, GRB 910601 and  GRB 910814 from approximately 20
keV to a few hundred MeV. 
Such broad band studies are, more or less, consistent with a 
continuation of the BATSE spectrum. 
In a few cases, very hard radiation has been observed to be emitted
late in, or even after, the main (lower energy) part of the burst.
For instance GRB 940217, a burst lasting for 160 s
as observed between 15 keV and 2 GeV, emitted 
GeV photons up to 1.5 hours after the trigger, with one photon having
an energy of 18 GeV (a significant detection; Hurley et~al. 1994). 

Barat et~al. (1998) present the spectra between 0.1 and 10 MeV
of the 20 most intense bursts observed by 
PHEBUS/{\it Granat} and report on the existence of a sharp break 
at typical energies of either around 1 MeV or around 2 MeV. 
They fit a 6 parameter function allowing for a second, high-energy
sharp break. The fit to the spectrum of GRB 900520a  is shown in the 
right-hand panel in Figure~\ref{fig25}.

Strohmayer et al. (1998) studied a number of GRBs observed by
{\it Ginga}, covering an energy band below the BATSE range ($\sim 2 -
400$ keV) and found a substantial number of bursts with breaks below
10 keV, i.e., below the observable range in BATSE. The authors 
propose that the observations are due to the existence of two
breaks in the GRB spectrum, one in the BATSE range and one below
this, close to 5 keV. This is also consistent with their 
finding that the X-ray spectra are often hard,
with  positive $\alpha$s, in 40 $\%$ of their studied sample.

The observed ratio of the energy emitted as X-rays (2-10 keV)
relative to the gamma-rays (50-300 keV) is often a few $\%$, but 
in some cases it can be substantially larger, giving an average of
24 $\%$, with a logarithmic average of 7 $\%$ for the 22 bursts 
studied by Strohmayer et~al. (1998).

\subsubsection{Soft Excess and Spectral Subclasses.}

Several early studies occasionally observed significant emission
in the X-ray range of GRBs ($\sim 2 - 20$ keV). This was done, for 
instance, by XMON/{\it P78-1} (Laros et~al. 1984) and
by WATCH/{\it GRANAT} (Castro-Tirado 1994). 
Preece et al. (1996c) studied
the  time-averaged spectra from 86 BATSE bursts in search of
a soft excess above the extrapolated low-energy power-law. 
They used
the  256 channel SD data and combined them with the lowest energy
SD discriminator channel, leading to a useful spectral coverage
from approximately 5 keV to 2 MeV, after making  a post-launch
calibration of the $5-20$ keV region. They searched for soft emission
below 20 keV and found this in 14 $\%$ of the cases. The enhancement was
$1.2 - 5.8$ times relating to the standard power-law model flux, exceeding
5 $\sigma$ in significance. Not a single case had a low-energy
flux deficit. In their study they also identified 4 cases with
a peak energy below 45 keV and with a $\beta \sim -2$.
For the cases which had a peak energy larger than 100 keV the averaged 
low-energy power-law had $\alpha \sim -1.0$, and for the cases with a
peak energy below 100 keV the averaged value was $\alpha \sim -0.3$.

Pendleton et~al. (1994) searched for spectral subclasses. They studied
206 bursts with the  LAD 
4 channel data and used a direct spectral inversion technique
to obtain the photon spectrum.
They found that the distribution of spectral states is 
broad and that there are no clear sub-classes, albeit a weak
clustering. The authors also found that  the 
peak fluence (the fluence in the 64 ms
interval with the highest count rate in channels 2 and 3) is
significantly harder than the total fluence (from all counts in the burst
interval) in the range 20-100 keV, which indicates that the
time-resolved spectra are flatter than the time-integrated
spectrum. This is also observed to be the case, for instance, 
by Ford et~al. (1995), Liang \& Kargatis (1996), and Crider et~al.
(1997, 1998a). The time-integrated 
spectra differ from the instantaneous spectra and it is of great
importance that this is considered when physical models are
tested. This is especially a problem in cases when the spectra evolve
markedly which makes the two spectra differ notably (see \S 
3 and, e.g., Crider et~al. 1997; Ryde \& Svensson 1999a).

Continuing the pursuit of spectral sub-classes,
Pendleton et~al. (1997) identified two distinct types of spectra.
They studied a sample of 882 bursts with the LAD 4 channel data.
The `not-high-energy-bursts' (NHEB) have a marked lack of fluence above
300 keV. The authors even study individual pulses within the burst and find
that 'high-energy-bursts' (HEB) can consist of the sum of
high-energy-pulses and
not-high-energy-pulses, while the NHEB  can consist only of 
not-high-energy-pulses. 
Bonnell \&  Norris (1999a, 1999b) argue that the NHE
class of GRB probably is due to a brightness bias in the observations. 

\section{Spectral Evolution}

\subsection{Time Evolution of the Spectral Hardness (Peak Energy)}

\begin{figure}
\centerline{\epsfig{file=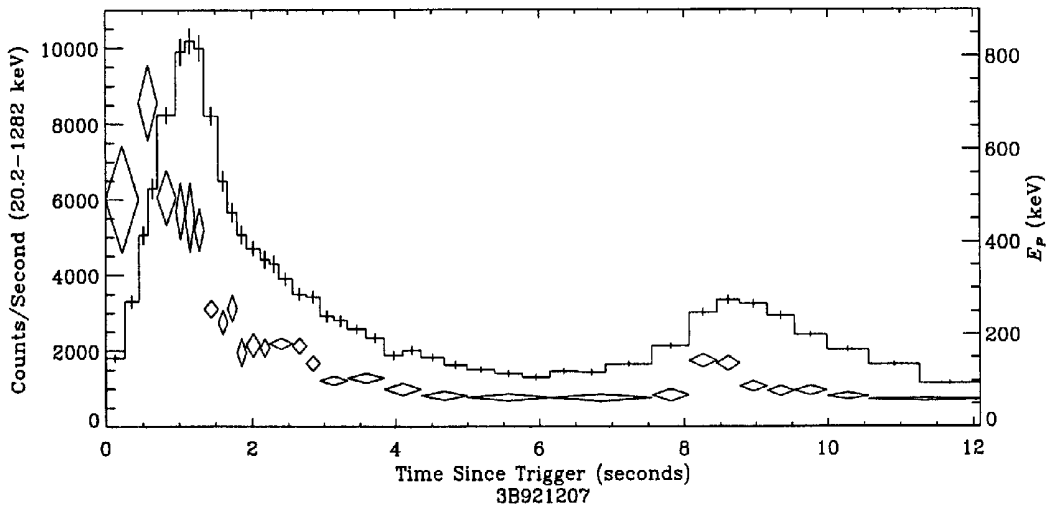,width=14cm,height=7cm}}
\caption{GRB 921207 (\#2083) in the study of Ford et~al. (1995).
The histogram represents the light curve in counts/s and
the diamonds represent the peak energy (in units of keV) measurements.
}
\label{fig4}
\end{figure}

It was early discovered that  the time-resolved (instantaneous) spectra 
in general soften with time (Mazets et~al. 1982; Teegarten et~al. 1982). 
In the survey of spectral evolution of BATSE bursts,
Ford et~al. (1995) found a number of common trends.
They studied 37 bursts using the SD data and concluded that the
peak energy rises with or slightly precedes intensity increases and
softens for the remainder of the pulse. 
They also found that successive pulses are usually 
softer, as well as that there is a general softening in time outside of 
the main pulses over the entire burst. 
Furthermore, bursts for which the bulk of the flux comes 
well after the trigger tend also to be softer. 
There were also a few bursts which did not show these behaviors. 
Figure~\ref{fig4} shows the result of the analysis of the strong
burst GRB 921207 (BATSE trigger \# 2083) in Ford et~al. (1995).
For the BATSE observations, the peak energy varies, in general, by a
factor of 5, with some cases reaching up to a factor of 15, over the burst.
Complex bursts have only a weak and slow time evolution.

The softening over the burst can, in some cases, be spectacular and have a 
complex behavior. Occasionally there is a correlation between the
spectral hardness and the intensity.
In most pulses, the hardness decays monotonically, creating the
{\it hard-to-soft} pulses, while in some the hardness tracks the
intensity, creating the {\it tracking} pulses, cf. \S 4.

Beside the general trends there are examples of bursts with very
diverse behaviors. 
For instance, GRB 980519 (\#6764), which was observed by {\it BeppoSAX} and
BATSE with a total energy range of $2 - 1900$ keV (in't Zand et~al. 1999)
exhibited a soft-to-hard-to-soft evolution.
The whole evolution seems to be connected, suggesting that
the soft initial phase is not a preburst X-ray activity, but may have 
a common origin with the main GRB emission. 

\subsection{The Evolution of the Spectral Shape}

A systematic investigation of the shape of the spectrum below the peak 
energy was made by Crider et al. (1997), using the 128 channel
LAD data.
They studied  the slope of the asymptotic low-energy power-law, in
terms of its index $\alpha$ in a sample of 79 bursts,
and found that $\alpha$ evolves in 58 $\%$ of the
cases.  Some bursts
exhibit substantial evolution in $\alpha$ over the burst.
Furthermore, they conclude that $\alpha$ follows the evolution
of the peak energy, both for hard-to-soft pulses and for tracking
pulses, albeit with less confidence for the latter result. The averaged
values of the power-law slope during the rise phase of the pulses are
significantly harder for the hard-to-soft pulses, with 40 $\%$ of
them having a positive averaged $\alpha$-value. 
The most extreme example of spectral {\it shape} evolution
is found in GRB 910927 (\#829),
in which the low-energy power-law index evolves from approximately $+1.6$
down to $-0.5$. The maximal value
of $\alpha$ is somewhat dependent on the analysis
and could actually be lower. However, it is beyond doubt that the $\alpha$s
can be large and values close to 0 are certain.
For the tracking
pulses the averaged value remains negative during the rise phase.

The spectral shape above the break energy, $E_{\rm pk}$, i.e., the 
high-energy power-law, does not change as much as has been observed
for the low-energy power-law.
Preece et~al. (1998a) studied the behavior of the high-energy
power-law in detail, using the 128 channel LAD data, which became useful 
after an in-orbit calibration. 126 bursts were studied: 122 of these 
 had a spectrum consistent with a power-law, and 
for the evolution of this power-law 34 $\%$
were inconsistent with a constant $\beta$.
The value of $\beta$ averaged over the burst has a narrow
distribution, $-2.12 \pm 0.30$.
There were a 
few events classified as super soft, having a $\beta < -3$,
cf. Pendleton (1997). 
100 events showed a hard-to-soft evolution and the averaged change of $\beta$, 
$\Delta \beta = -0.37 \pm 0.52$. Furthermore, it was found that the
behavior of $\beta$ is independent of the rest of the spectral
evolution.

\section{Connection between the Spectral and the Light Curve 
(Intensity) Behavior}

\begin{figure}
\centerline{\epsfig{file=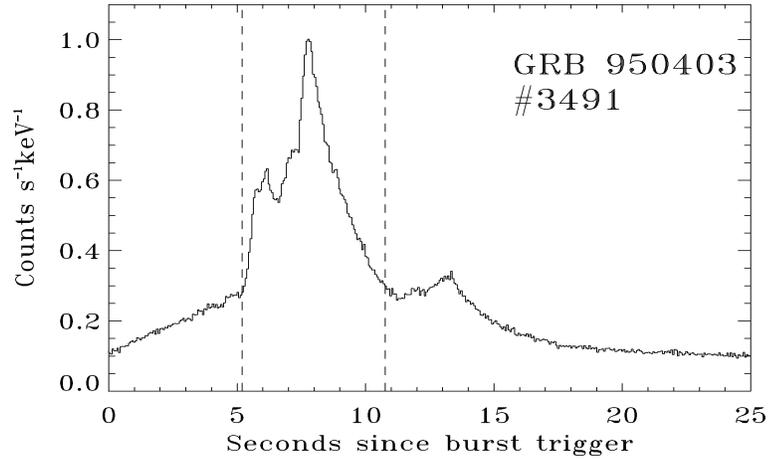,width=10cm,height=6cm}}
\caption{Light curve of GRB 950403 (\#3491), observed by the LAD
(4 channel data; all four energy channels). The dashed
lines indicate the interval displayed in Figure~\ref{fig47}.}
\label{fig43}
\end{figure}

\begin{figure}
\centerline{\epsfig{file=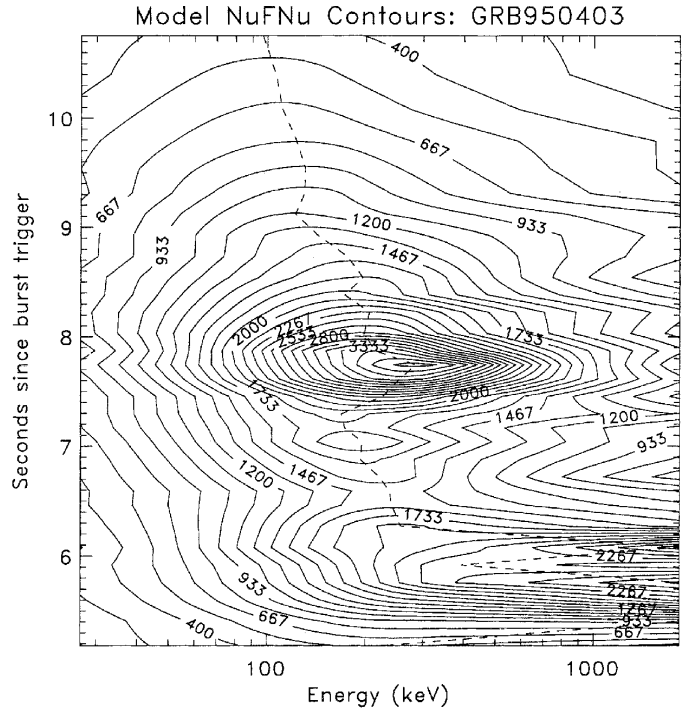,width=10cm,height=10cm}}
\caption{Topological map of the GRB-cube for 
GRB 950403 (cf. Figure~\ref{fig43}). The $E F_{\rm E} (E,t)$ 
(keV cm$^{-2}$ s$^{-1}$) 
contours of the fitted `GRB-function' are
displayed. The dashed line follows the time evolution of the peak
energy. Adapted from Preece et~al. (1996b).
}
\label{fig47}
\end{figure}

The spectral evolution is more or less coupled with the intensity of 
the burst.
By studying narrow time bins of the light curve,
the instantaneous spectra can be studied, giving
information on the temporal behavior of the spectrum over the burst.
Correspondingly, the light curves from different spectral
energy  channels show how the intensity of different energies 
compare with each other.  
The burst evolution can be described as taking place in an imaginary
cube, having the spectral energy and the time axes in the x-y-plane and the
intensity on the z-axis: `the GRB-cube'. The full evolution can then 
be illustrated as contour plots of the intensity on the energy-time
plane. An example of such a plot is given in Figures~\ref{fig47} for 
GRB 950403 (\#3491), whose light curve, from all the 4 LAD channels, is
shown in Figure~\ref{fig43}. The contours are from the
fitted GRB-function, and the evolution of the peak energy is indicated. 

To describe this evolution and to systematize the observations
in order to see the general trends,
empirical relations between observables have been sought. 
The physical reason for these correlations is then explored.
The main observables studied are the instantaneous 
photon (or energy) flux, the
spectral hardness characterized by the peak  (break) energy or
equivalently the temperature or color, the fluence and the total flux, the
spectral shape parameters (e.g., the power-law indices), and the duration
of the pulse and burst.

\subsection{Quantitative Correlations}

\subsubsection{Hardness-Intensity Correlations (HIC).}

The relation between the intensity and the hardness  has been 
well investigated and it has been shown that there is no ubiquitous
trend of spectral evolution that can characterize all bursts; 
several types of behavior exist.
Firstly, Norris et~al. (1986) found that the most common trend of spectral
evolution is a  hard-to-soft behavior over a pulse,
with the hardness decreasing monotonically as the flux rises and
falls. They studied 10 bursts observed by the {\it Solar Maximum 
Mission} satellite.
This behavior was also seen to be the most common trend by  
Kargatis et~al. (1994), who studied 16 SIGNE/{\it Venera} bursts.
There are also a few cases which exhibit 
soft-to-hard and even soft-to-hard-to-soft evolution.
In a study by Band (1997), 209 BATSE bursts were studied 
with the LAD discriminator rates giving high time resolution. 
The spectral evolution was studied
through auto- and cross-correlation between light curves from the four LAD
channels. Most of the  bursts in the sample showed a hard-to-soft
behavior.

Secondly, there is a tracking behavior between the intensity and the hardness,
first noted by Golenetskii et~al. (1983). 
Kargatis et~al. (1994), confirmed the
existence of such a HIC, even though it was less common than the
hard-to-soft trend. However, in the decay phase of hard-to-soft 
pulses the HIC is often seen. Kargatis et al.  (1995)  found the
hardness-intensity correlation 
in 28 pulse decays in 15 out of 26 GRBs with prominent pulses.
Ryde \& Svensson (1999b) also studied the HIC for the decay phases 
of a number of strong burst pulses.

Thirdly, there are bursts that do not exhibit any correlation at all having a
chaotic behavior. Indeed, the main conclusion drawn by
Jourdain (1990), who studied several
bursts observed by the {\it APEX} experiment, and Laros et~al. (1985), 
who studied a few {\it Pioneer Venus Orbiter} bursts, is
that there does not exist any correlation between the spectral
evolution and the time history in their samples of GRBs.
Over the whole GRB, there often does not exist any pure  correlation,
even though the tracks in the hardness-intensity plane are confined
to an area from hard and intense to soft and weak, indicating an
overall trend with increasing luminosity with hardness. 
(Kargatis et~al. 1994). A
chaotic behavior in the plane 
may be a result of the superposition of several short
hard-to-soft pulses that cannot be resolved.

Several of the different types of trends can also be seen in a single
GRB (e.g., Hurley et~al. 1992).
The variety of behaviors is also manifested in Band et~al. (1993)
and Ford et~al. (1995). 
Bhat et~al. (1994) studied 19 time structures,
which have a FRED-like shape with a short rise time
($< 4$ s), and found that most had a good correlation between the
hardness and the intensity.

\begin{figure}
\centerline{\epsfig{file=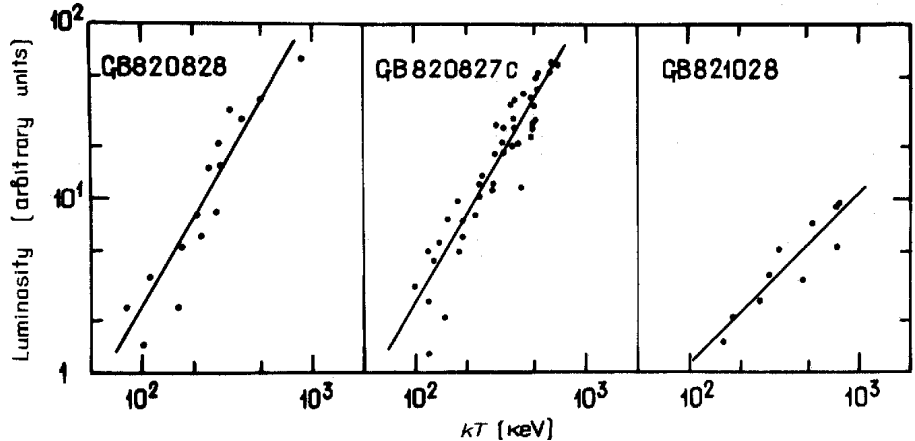,width=12cm,height=6cm}}
\caption{Fits to the hardness-intensity correlation for three GRBs
observed by KONUS/{\it Venera}.  The temperature is used as a measure
of the hardness. Adapted from Golenetskii et~al.  (1983).
Used with permission from Nature.
}
\label{fig85}
\end{figure}

The tracking behavior has been described quantitatively.
Golenetskii et~al. (1983) 
found the power-law relation between the instantaneous luminosity
($\propto$ the energy flux) and the peak energy
\begin{equation}
L \propto (k T)^\gamma,
\end{equation}
where the peak of the spectrum was quantified as the temperature, $T$, in the
thermal bremsstrahlung model ($k$ is the Boltzmann's constant). 
The power-law index (the correlation index), $\gamma$, was found to 
have a typical value of $1.5-1.7$. Figure~\ref{fig85} 
shows fits to the hardness-intensity correlation
for three GRBs discussed in the original paper by Golenetskii et~al. 
(1983).
This analysis was criticized by several workers, including 
Laros et~al. (1985), 
Norris et~al. (1986), and Kargatis (1994).
It was speculated that the correlation could possibly
be an artifact from the way the temperature was derived from the 
two-channel count rates. Furthermore,
Golenetskii et~al. (1983) excluded the hard initial phase of the
bursts. Ford et~al. (1995) suggested that the low time-resolution
may result in the initial, non-tracking, hard behavior being missed.

However, Kargatis et~al. (1994)
confirmed the existence of the Golenetskii et~al. (1983) correlation
in approximately half of their cases.
The spread was substantially wider $ \gamma =
2.2 \pm 1$. 
In the Kargatis et~al. (1995) study, in which 
the decay phase of a number of prominent pulses were examined,
it was found that the distribution of 
the correlation index peaks at 1.7 and has a substantial spread.
Bhat et~al. (1994) found a corresponding spread in the
HIC index.

In the study of the {\it Ginga} data, Strohmayer et~al. (1998)
investigated the evolution of the peak energy versus energy flux
and found the power-law
correlation to be valid here too, with, for instance, $\gamma \sim 3$ 
for GRB 890929 (in the {\it Ginga} energy range).

\subsubsection{Hardness-Fluence Correlation (HFC).}

\begin{figure}
\centerline{\epsfig{file=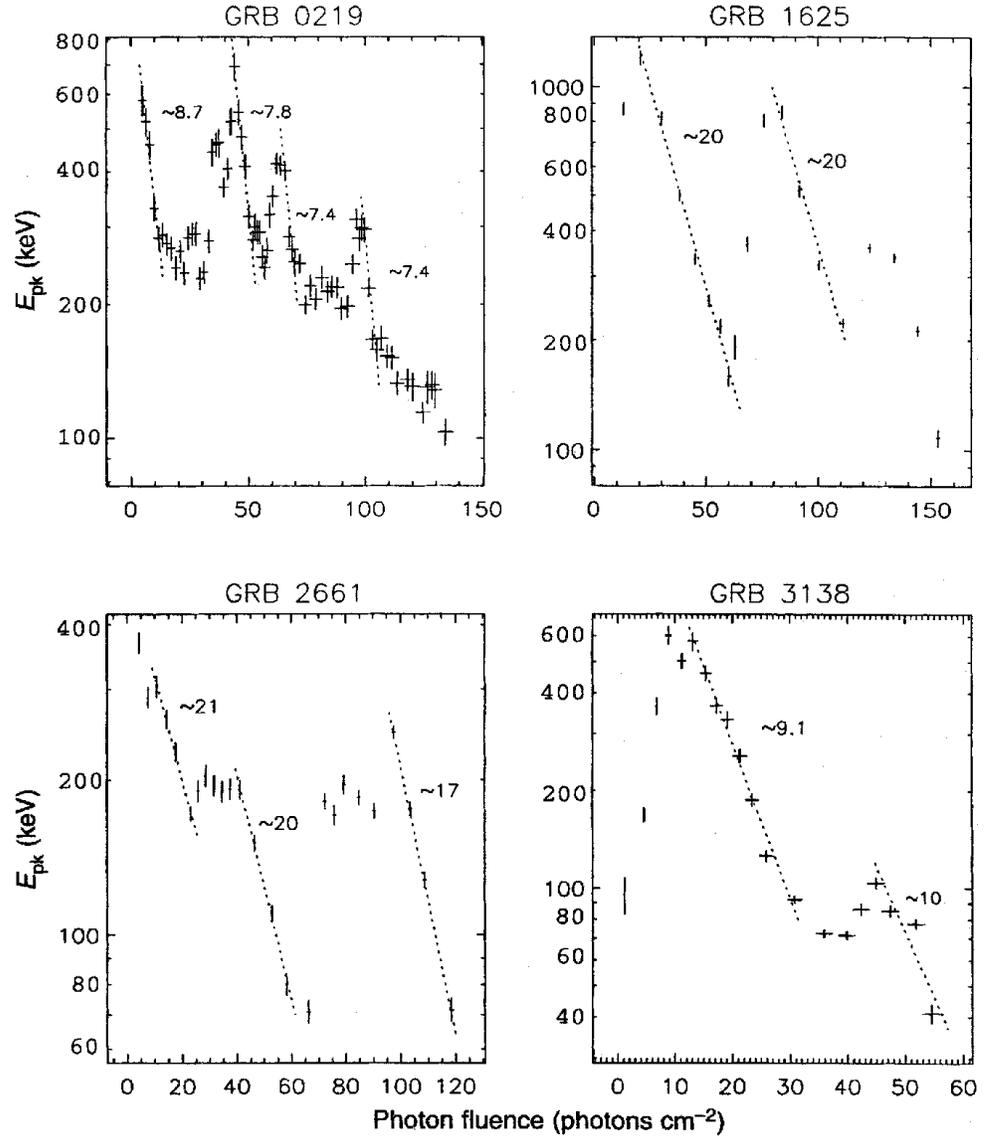,width=14cm,height=16cm}}
\caption{Fits to the hardness-photon fluence correlation of four GRBs
observed by BATSE; GRB 910522 (\#219), GRB 920525 (\#1625),
GRB 931126 (\#2661) and GRB 940826 (\#3138).
The bursts have several pulses with almost
invariant decay constants $\Phi_{\rm 0}$. From Liang \& Kargatis (1996).
Used with permission from Nature. 
}
\label{fig9}
\end{figure}

As mentioned above, the correlation between the hardness and the 
flux (luminosity) over the entire burst or even a pulse does not always
show any clear correlation. However, by studying the relation between the
hardness and the running time-integral of the flux, the fluence, a
clear correlation is often revealed over the entire pulse.
Liang \& Kargatis (1996) consequently 
found an empirical 
relation defining how the instantaneous
spectrum evolves as a function of photon fluence, $\Phi (t) = 
\int^t N(t')\,dt'$. They found that the power peak energy
of the time-resolved spectra of a single 
pulse decays exponentially  as a function of $\Phi (t)$, i.e.,
\begin{equation}
E_{\rm pk} (t) = E_{\rm pk,max} e ^{-\Phi (t) / \Phi_{\rm 0}},
\end{equation} 
\noindent
where $E_{\rm pk,max}$ is the maximum value of $E_{\rm pk}$ 
within the pulse and $\Phi _{\rm 0}$
is the exponential decay constant. The photon fluence is the photon flux
integrated from the time of $E_{\rm pk,max}$. 
Figure~\ref{fig9} shows examples of fitted correlations from the
original work.
The authors found that 35 of the 37 pulses in the 
study were consistent or marginally consistent with the 
relation. Furthermore, they concluded
that the decay constant is constant from pulse to pulse within
a GRB. This view was, however, changed by Crider et~al. (1998a) who 
dismissed the apparent constancy as consistent with drawing values
out of a narrow statistical distribution of $\Phi_{\rm 0}$,
which they  found 
to be log-normal with a mean of ${\rm lg}\, \Phi_{\rm 0} = 
1.75 \pm 0.07$ and a FWHM of $\Delta {\rm lg}\, \Phi_{\rm 0} = 1.0 \pm
0.1$. 
This result is probably affected by selection effects.
 They expanded the study to include  41 pulses within
26 bursts, by using the algorithm introduced by Norris
et~al. (1996), to identify pulses. Another approach was also introduced,
where they used the energy fluence instead of the photon fluence.
The two approaches are very similar and do not fundamentally
change the observed trends of the decay.
These results confirm the correlation and extend
the number of pulses in which the correlation is found.
The relation between the two approaches is
discussed in \S 5.3. 

\subsubsection{Other Correlations.}

A few other correlations within individual GRBs should
also be mentioned. 
Norris et~al. (1996) introduced the asymmetry/width/softness 
paradigm for pulses, in which the quantities are correlated.
They only detect, however, a slight trend that more symmetric, narrower pulses 
are harder. Furthermore, Kouveliotou et~al. (1992) reported on a trend that 
pulses with a short rise time are harder in single-pulse events.
Using {\it PVO} bursts, Lochner (1992) noted a negative correlation 
between hardness and time between pulses.

The spectral evolution in gamma-ray color-color diagrams,
i.e, the correlation between hardness ratios have been studied. 
Kouveliotou et~al. (1993) could classify about half of the 30
bursts they studied into three types of behavior, crescent, island-like and
flat. They did not find any striking correlation between the temporal
profile of the bursts and the shape in its color-color diagram.

In the study of Lee et~al. (1998), the 2500 individual channel 
pulse structures analyzed also confirmed the general behaviors that pulses
are narrower and occur earlier at high energies. There is also a
negative correlation between the peak
flux and the pulse width, and between the pulse fluence and 
the pulse duration, within a burst. 
Petrosian et~al. (1999) discuss this and
note that these correlations are the same as 
the ones attributed to cosmological effects found in ensembles of
bursts. They therefore draw the conclusion that these ensemble correlations 
cannot be cosmological signatures alone, but must arise from 
the intrinsic properties of the GRBs.
 

\subsection{Quantitative Temporal Descriptions}

A few attempts have been made at quantitatively describing the 
temporal intensity-spectral evolution within a GRB pulse.
Fenimore et~al. (1995) studied how the width of a GRB pulse changes
as a function of spectral energy and  found that it 
scales as $E^{-0.4}$. This result was found both by using the
autocorrelation function for GRBs and by using the width of the
average pulse profiles for the four BATSE channels (see in 't Zand \& 
Fenimore 1996 for a discussion on the autocorrelation function).
The behavior was also observed in the whole band from 1.5 to 700 keV
by {\it BeppoSAX} (Piro et~al. 1997), which suggests that the emission 
mechanism is the same from soft X-rays to gamma-rays.
Following the notation (with some modifications) of
Fenimore \& Bloom (1995) the light curve in each of the 4 BATSE
channels ($k=1,2,3,4$) can be described as
\begin{equation}
H_{k}(t)= \int^\infty _0 R_{k}(E) A(E,t) N_{\rm E}(E) dE
\end{equation}
\noindent
assuming that the time structure can be separated from the spectral
shape, $N_E(E,t) = A(E,t) N_{\rm E}(E)$.
$R_{k}(E)$ is the effective area of the detector for each channel. 
The scaling factor, $A(E,t)$ was modeled by Fenimore \& Bloom (1995) as 
$A(E,t) = \exp[-t/\tau(E)]$ with $\tau(E)= S_1(E/100)^{S_2}$, $E$ is 
measured in keV and for the decay phase of the pulse, typically,
$S_1 = 0.45$ and
$S_2 = -0.39$ and for the rise phase, $S_1 = 0.22$ and $S_2 = -0.40$.

Neither the hardness-intensity correlation (Golenetskii et~al. 1983)
nor the hardness-fluence correlation (Liang \& Kargatis 1996) 
include the time dependence of the
spectral evolution. However, combined they do, as 
the fluence in the time integral of the flux.
This was used by Ryde \& Svensson (1999b) to synthesize and find a
compact and quantitative description of the time evolution of the decay
phase of a GRB pulse. This description is for the intensity-time plane
 of the GRB-cube, 
rather than for the single-spectral-channel light curves, which 
have been studied extensively in connection with their dependence with
energy. Ryde \& Svensson (1999b) identify a subgroup of GRB pulses, 
for which
the two empirical relations are valid and show that for these the
decay phase of the pulse should follow a power-law.
For the decay phase the HFC becomes
\begin{equation}
E_{\rm pk} (t) = E_{\rm pk,0} e ^{-\Phi (t) / \Phi_{\rm 0}}  ,
\end{equation}
where $E_{\rm pk,0}$ is the peak energy at the start of the
decay. Note that $E_{\rm pk,max}$ could be even larger, for instance
for hard-to-soft bursts.
For a moderate spectral shape evolution the HIC can be rewritten,
using the photon flux, as, in that case, it  holds that 
$E_{\rm pk}(t) N(t) \sim F(t)$.
For the decay phase we then have 
\noindent
\begin{equation}
\Epk(t) = E_{\rm pk,0}\left[\frac{N(t)}{\N0}\right]^{\delta},
\end{equation}
\noindent
where $N_{\rm 0}$ is the photon flux at the same time as 
$\Epk = E_{\rm pk,0}$, i.e., at the beginning of the decay phase.
The correlation index, $\delta$, corresponds approximately 
to $1/(\gamma-1)$, where
$\gamma$ is the index used by Golenetskii et~al. (1983).
These two relations, given by equations (6) and (7), fully describe the
evolution and especially the time dependence.
If these two relations are fulfilled the time evolution can be described
by a vector function $\Gt = (N(t),\Epk (t))$ given by
\begin{eqnarray}
N(t) & = & \frac{\N0}{(1 + t/\tau)};\\
\Epk(t) & = & \frac{E_{\rm pk,0}}{(1 + t/\tau)^\delta},
\end{eqnarray}
\noindent
where the initial value is ${\bf G}(0) = ( N_{\rm 0}, E_{\rm pk,0} )$ and 
the number of additional parameters is
limited to two, the time constant $\tau$ $\left[ N(t=\tau) = N_{\rm
0}/2 \right]$ and  the HIC index 
$\delta$. Note that the origin of the time variable, $t$, is
at the time of the intensity peak.
The peak energy has a similar dependence as the intensity,
differing only by the power
law index $\delta$.
The exponential decay constant of the HFC is given by $\P0 \equiv \N0 \tau /
\delta $, and thus the characteristic time scale of the decay, the  time
constant, $\tau \equiv \delta \P0/\N0$.

\begin{figure}
\centerline{\epsfig{file=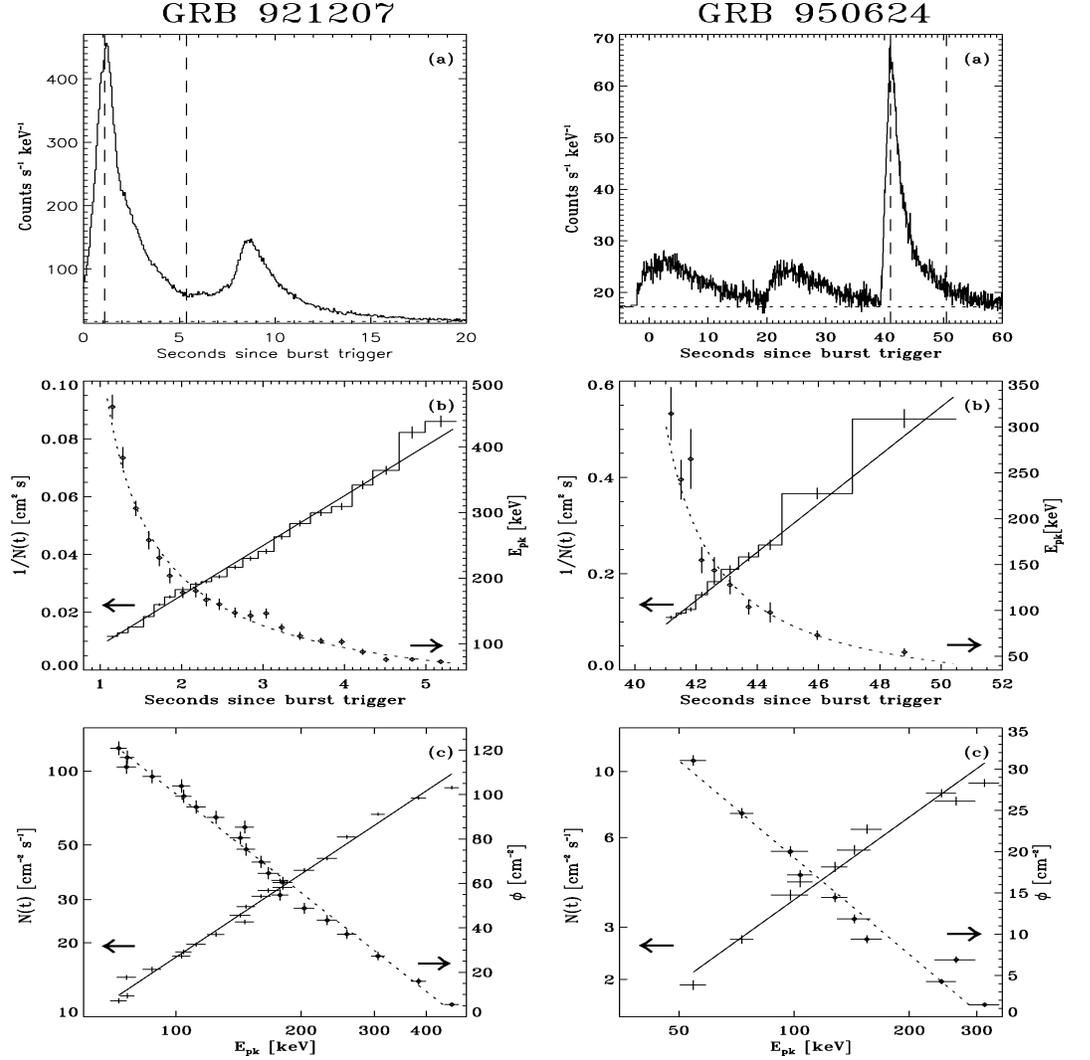,width=14cm,height=14cm}}
\caption{Panels (a): BATSE-LAD light curves of GRB 921207 (\#2083) and
GRB 950624
(\#3648). The decay phases of the dominant pulses are indicated by the 
verticle dashed lines and are examined in detail in panels b and c.
Panels (b): Fits of the equations (8) and (9)
to the indicated decay phases. Panels (c):
Fits of the empirical relations (6) and (7) to the indicated decay
phases. These results suggest that these two pulses belong to the
subgroup of GRB pulses for which the decay phase follows a power-law.
The parameters found from the fits to the empirical relations in panel
c (beside the initial conditions $N_{\rm 0}, E_{\rm pk,0}$),
i.e., $\delta$ and $\Phi_{\rm 0}$ are consistent with the
parameter values found from fitting the decay phase of the pulse;
$\tau \equiv \Phi_{\rm 0} \delta / N_{\rm 0}$.
See Ryde \& Svensson (1999b) for details.}
\label{fig10}
\end{figure}

The formulation, given by equations (8) and (9), is a condensate
of the HIC and the HFC, which have
been proven to be valid in several cases. Ryde \& Svensson (1999b)
studied a number GRB pulses in this context, fitting both the original 
correlations, as well as the new equivalent formulation.
The fits of the behavior of the decay phases of two such strong 
pulses are shown in Figure~\ref{fig10}.
This shows, among other things, that if the HIC and the HFC are valid
the decay part of the light curve (intensity) should follow the
power-law $N(t) \propto (1+t/\tau) ^{-1}$.
This behavior cannot persist too long as the integrated flux
(the fluence) has a divergent behavior 
$\Phi (t) = N_{\rm 0} \tau {\rm ln}(1+t/\tau)$. 
The decay of the intensity must thus change into a more rapid one,
such as an exponential, or possibly be turned off completely.

\subsection{Relation Between the Time-Integrated and Time-Resolved
Spectra}

The hardness-fluence correlation 
gives us the possibility to understand in what way the
instantaneous and the time-integrated spectra are related over a pulse.
What time-integrated spectrum  does this relation 
give rise to? I.e., what does the spectrum on the intensity-energy
plane of the GRB-cube look like?
The exponential decay of the peak energy with fluence means that a 
linear increase in fluence by equal steps of
$\Phi_0$ photons cm$^{-2}$ corresponds to a decrease of
$\ln E_{\rm pk}$ in equal logarithmic steps.
As the instantaneous spectra are, roughly, dominated by the 
$N \sim \Phi_0$ photons cm$^{-2}$ in
a logarithmic interval around the peak energy,  
$dN / d\ln E$
$ = E dN / dE \equiv E N_{\rm E}(E)$ is a constant $= \Phi_0$.
In other words, the time-integrated,
specific flux spectrum is a constant function of energy,
and thus the time-integrated photon spectrum, 
$N _{\rm E}(E)$ of a single pulse has a power-law slope
of $-1$. This is a 
direct result of the specific evolution defined by the
hardness-fluence correlation. I.e.,  the spectral 
shape is a result of the {\it exponential} decay 
of the peak energy versus photon fluence.
This spectral shape is reminiscent of the optically-thin thermal 
bremsstrahlung spectrum.
This was studied in detail by Ryde \& Svensson (1999a), who showed
analytically how the time evolution of the 
instantaneous spectra is related to the resulting time-integrated
spectrum.  They studied mainly the spectra of 
single FRED pulses and showed that the
exponential decay of the peak energy with photon fluence, indeed, does 
lead to a general, low-energy slope, normalized by the decay constant 
$\Phi_{\rm 0}$ and having the underlying $E^{-1}$ behavior.
This general result is 
affected by the finite range over which the peak energy evolves; the
less it evolves the more the spectrum is affected. The way the
spectrum is affected can be found analytically, leading to a function 
that can be used to fit the time-integrated spectra, and having
parameters describing the instantaneous spectra.
From the fit to the time-integrated spectrum one can 
deduce information about the instantaneous spectra, which is of
interest as these carry more direct physical information. 
This is not the case for the time-integrated spectra,
as they are merely the result of the exponential decay of the peak
energy.

\section{Analysis Methodology}

\subsection{Data Analysis}

The observed distribution of the spectral parameters
could very well be different from the parent distribution
due to observational biases, such as truncation of the data
set resulting from the trigger procedure.
For instance,
is this the case for the narrow range of peak energies found by 
the BATSE instrument?  Cohen et~al. (1998) argued  that
the detection efficiency of the BATSE could lead to an unreal paucity
of hard bursts and they suggest that there could exist a large, unobserved
population of hard (MeV) bursts. If the luminosity at the peak
energies is represented by  a standard candle,
high peak energies will result in fewer photons, letting fewer pass the 
trigger. Very low peak energies will correspondingly affect the
triggered fraction, as the spectra would have their cut-offs below the
detectable range. 
This is also noted in Lloyd \& Petrosian (1999) and 
Petrosian et~al. (1999) who show that the effects
of selection biases and data truncations are to produce observed
distributions that are narrower than the parent distributions. They
also present methods to properly account for this.

Furthermore, are the assigned values of the spectral
characteristics, mainly the peak energy, $E_{\rm pk}$, and 
the slope of the asymptotic, low-energy power-law, $\alpha$,
correctly measured? 
As it is the asymptotic value of the power-law that is measured, problems 
can arise the closer the energy break gets to the low-energy 
cut-off of the energy window of the detector (generally 
15-25 keV for BATSE data). The fitting can lead to a wrong
value being ascribed to $\alpha$, and consequently also to $E_{\rm
pk}$, as the exponential turnover (curvature) is not modeled correctly.
This is also evident in the comparison of the results from fitting 
a sharply broken power-law to the data, instead. This function does not take
the exponential turnover into account and therefore gives a steeper
(lower $\alpha$) power-law compared to the asymptotic value.
 The errors in the fitted
$\alpha$-parameter also increase as the available energy range,
for fitting, decreases.
Preece et~al. (1998b) use the power-law tangent  to the `GRB' function
at some  chosen low energy (e.g. 25 keV)  as the upper bound of the
low-energy power-law behavior within the observed energy window. 
This value is, however, always smaller than the asymptotic value.
In addition, the values assigned to the fitted parameters can 
be erroneous due to the existence of any
soft component, any previous pulses adding soft photons 
to the spectrum, or any completely unresolved pulses
with lower break energies. These 
issues could affect the estimated fraction of pulses over which the 
spectral shape actually  changes. 
Kargatis et~al. (1995) and Liang \& Kargatis
(1996) freeze the power-law values to their average values during the
burst, which reduces the spectral variations into merely the hardness
variation. Correspondingly, the measured value of $\beta$ is
sensitive to the amount of high-energy signal available for its
determination. The power peak, in some cases, does not lie within the
BATSE band, e.g., when $\beta > -2$. For a detailed discussion
on these issues see Preece (1999).

To be able to study the spectral evolution on finer time scales a
coarser hardness measure in needed.  Often the hardness ratio
between different energy bands is used.
Bhat et~al. (1994) compared the analysis of the spectral evolution
with the two different hardness definitions, conventional spectral 
fitting and hardness ratios, and found that the two were 
consistent.

Furthermore, does the BATSE spectrum, between 20 keV and 1900 keV, 
represent a correct measure of the bolometric flux? 
The energy spectrum often peaks within the BATSE band and thus 
it should be a good measure of the bolometric {\it energy} flux.
Such  considerations made Crider et al. (1999)
prefer to study the spectral evolution in terms of energy flux rather
than photon flux, cf. \S 5.3.

Moreover, the evolution of the spectral characteristics and the
correlations between them could also be affected by various
limitations in the observations.
Schaefer (1993) discusses methodological problems in connection with
the study of the HIC and expresses concern with  the fitting technique
used (Isobe 1990). He emphasizes the importance of having high spectral 
resolution rather than time resolution, so as not to introduce  
artificial correlations.
To do spectral analysis correctly it has been shown that the 
signal-to-noise ratio should be of the order of $45$ in
the BATSE range (Preece et~al. 1998a).

\subsection{Detailed Spectral Modeling}

To model the time-integrated photon count spectra, from which 
the background has been subtracted, the `GRB' function (Band
et~al. 1993) is the most commonly used, cf. \S 2.2.
It is
a purely empirical model described by 4 parameters. Besides the
normalization, the two power-laws and the break energy are fitted.
Earlier studies used the 
`optically-thin thermal bremsstrahlung' (OTTB)
spectrum, ${N} _{\rm E} (E)\propto E^{-1} {\rm exp} ({-E/kT})$
and the 'thermal synchrotron' (TS) spectrum from an optically-thin,
mildly relativistic, thermal plasma in a magnetic field $B$; 
${N} _{\rm E} (E) \propto  {\rm exp} \left[ {-(E/E_{\rm
c})^{1/3}}\right]$, where $T$ is the temperature and 
$E_{\rm c}$ is the critical frequency proportional to $B T^2$ (Liang 1982).
The empirical models are often 
used only to determine the general shape, e.g., the hardness, and not 
to determine physical characteristics of the source, like a
temperature.
The value assigned to the break of the spectrum depends on the model
used and can thus affect the correlations sought for. The break is
determined from the fit to the overall continuum shape which is
modeled in different ways. 
This was noted, for
instance, by  Schaefer et~al. (1992) and Kargatis et~al. (1994).
In the latter study the authors used the OTTB and TS models and found that
many cases gave consistent results but that there were cases for which rather
different values were obtained. 
Another such study was done by Ryde (1999), where 10
GRB spectra were fitted with  three different models: a sharply
broken power-law, the `GRB' model, both with 4 parameters, and the
smoothly broken power-law,
described by a broken power-law, smoothly and evenly 
connected through a hyperbolic function with a total of 5 parameters.
The extra free parameter describes the width of the transition
region (see, e.g., Preece et~al. 1996a; Ryde 1999).
In some cases, the peak energies
attributed to the data differ considerably. As bursts can have an actual
curvature that is sharper
than the one given by the fixed exponential curvature in the `GRB'
function, which is determined solely by $E_{\rm 0}$, the resulting
fits may differ.
Adding extra parameters to the model function is meaningful only
if the data are good enough to enable a constraining of the parameters.

\subsection{Energy or Photon Flux?}

The intensity can be characterized either in terms
of the photon/count flux or of the energy flux. For instance,
Golenetskii et~al.  (1983) studied   the HIC based on the energy flux 
(luminosity),
while Bhat et~al. (1994) studied the correlation using the detector
count flux instead, allowing for higher time resolution
by using hardness ratios to characterize  the spectrum. The 
studies arrive at similar conclusions for the correlation. 
Liang \& Kargatis (1996) choose to study the photon flux in their search of
the HFC. This was motivated by the fact that the decay of the
hardness versus energy fluence was more difficult to establish because 
the deconvolved energy flux has larger statistical errors. Liang \&
Kargatis (1996) discovered the HFC using the photon fluence. In a
larger study of the HFC, Crider et ~al. (1999) used the energy flux
instead. More detailed spectral fitting can reduce the statistical
errors
in the energy fluence. The authors also found the HFC in the energy flux fits 
for all the 41 cases they studied. In parallel, they also studied the
decay versus photon 
fluence suggested originally, and confirmed the discovery. 
The reason they
prefer the energy fluence over the photon fluence is their argument
that the energy fluence represents a more physical quantity and that
it is a better measure of the  bolometric flux. However, they
point out that the two approaches do not represent 
fundamentally different trends.
Ryde \& Svensson (1999b) used the photon flux both
for the HIC and for the HFC in their synthesis of the  spectral 
evolution of a GRB pulse and showed the relations to hold. What then is
the difference and when is a difference between the approaches expected 
to be seen? Can we determine which correlation is the most
fundamental, i.e., the one always valid and not merely a consequence
of the other?

The decay tested for is either the exponential decay of the peak energy versus
photon fluence,
\begin{equation}
E_{\rm pk}(t) = E_{\rm pk,max} e ^{-\Phi(t)/\P0},
\end{equation}
\noindent  
or the linear decay of the peak energy versus energy fluence, ${\cal
E}(t)$,
\begin{equation}
E_{\rm pk}(t) =   E_{\rm pk,max} - \frac{1}{\P0} 
\int _0 ^t F(t')\,dt' =  E_{\rm pk,max} - \frac{1}{\P0}{\cal E}(t).
\end{equation}
\noindent
Differentiating equation (10) gives the decay rate of the peak as
\begin{equation}
-\frac{dE_{\rm pk}(t)}{dt} = \frac{E_{\rm pk}(t)}{\P0}N(t) \sim 
\frac{F(t)}{\P0}.
\end{equation}
\noindent
The last step in equation (12) is generally only approximately 
true. 
The equivalence between the exponential decay in photon fluence,
equation (10), and the linear decay in energy fluence, equation (11),
depends on the validity of this approximation.

The energy flux
\begin{equation}
F(t) = \int_0 ^\infty E \cdot N_{\rm E}(E,t)\,dE =
N(t)\int_0^\infty E \cdot f_{\rm E}(E,t)\,dE \equiv N(t) \langle E \rangle,
\end{equation}
\noindent
where $\langle E \rangle$ is the flux-weighted, averaged energy, i.e.,
the mean  energy,
and $f_{\rm E}$ is the normalized spectrum.
The assumption that the two decays are the same is
equivalent to that $\langle E \rangle = E_{\rm pk}$. This is exactly
the case in the 
often illustrative Dirac $\delta-$function approximation of the
spectrum. It  is also 
the case, when the spectral shape, 
i.e., the function $f_{\rm E}(E,t)$, is symmetric around the peak
energy. The approximation is better the
more peaked the logarithmic $E N_{\rm E}$-spectrum is. 

A complication to the
discussion stems from the fact that $\langle E \rangle$ can shift as 
$f_{\rm E}(E,t)$ changes,
i.e., when the spectral shape varies, for instance, with an evolving 
$\alpha$.  
Furthermore, when the spectral shape changes, the relation between the
power peak, $E_{\rm pk}$, and the photon number peak, $E_{\rm p}$,
varies, as $E_{\rm pk} = 
(2+\alpha)/(1+\alpha) E_{\rm p}$. For which measured peak energy 
does  the relations hold the best? Ryde \& Svensson (1999a) argue 
that the $E_{\rm p}$ is the important measure.
General uncertainty in the measurement of the peak energy has already
been discussed above.
These issues add to the fact that the data cannot clearly
demonstrate which relation is the correct one. In other words,  
we cannot make any
conclusive statement based on information gleaned from the data alone.


\section{Discussion}

Gamma-ray bursts are at cosmological distances,
 as indicated by recent observations 
of the afterglow, giving high redshift values 
(e.g., Costa et~al. 1997; Metzger et~al. 1997).
One plausible origin of
the huge energy release needed, is from a dissipative, relativistically
expanding fireball (blast-wave), or, equivalently, a propagating jet
with low baryonic contamination (e.g., M\'esz\'aros \& Rees 1997). 
The motivation for this scenario is based on the requirement that the observed
amount of energy must be injected inside a very small volume, given by 
the characteristic time scales of GRBs. 
The  photon energy
densities imply that the radiation is super-Eddington 
(by orders of magnitude; particularly if the radiation is 
emitted isotropically) and lead to the creation of an optically
thick, dense radiation and
electron-positron-pair fluid, expanding under its own pressure  
and cooling adiabatically.
The observed radiation can thus not come from
the surface of the central energy source.
Initially the fireball is thermal and converts the radiation energy
into bulk kinetic energy. The thermal emission from the 
fireball will not be visible in the
gamma-ray band, but may be visible at lower energies.
It then becomes optically-thin and the kinetic energy of the wind will
be tapped by a dissipation mechanism, such as shocks or 
magnetohydrodynamic (MHD) turbulence,
converting it into internal energy and accelerating relativistic
particles. The shocks could occur as the fireball
crashes into the circum-burst, low-density environment or as different
shells, with different Lorentz factors, within the fireball catch 
up with each other.
The general hard-to-soft evolution of the burst spectra
could then be explained by the expansion and deceleration of the
blast-wave and/or by the decline in the averaged
available energy, as more particles reach the shocks.
The pair plasma wind has to be highly relativistic with
Lorentz factors of $\Gamma \sim 10^2 - 10^3$ to avoid photon-photon
degradation through pair-production, as high-energy photons are observed.
 This implies that the baryonic pollution of the radiation
fields cannot be very high ('clean fireball'). Lower Lorentz factors
are, however, possible but then the production of the high-energy
radiation cannot be directly connected to the lower energy gamma-rays.

The primary source and signatures of the underlying mechanism
is enshrouded by the optically-thick pair plasma at the
beginning of the life of the fireball and details are washed out 
and cannot be seen directly in the observations.
The observed radiation from this 
initial phase is from and outside of the pair photosphere. Thus the
nature of the primary energy release will not greatly affect the
resulting expanding fireball.
Models including stellar-mass, compact objects, such as a merging 
neutron star and a black hole, meet the requirements of
occurrence and energetics. 
This event can either be a single, short-lived, catastrophic event,
producing a single fireball, or result in a recurrent central engine
capable of producing several shells. The latter could, for instance, be
a long-lived accretion system with the debris of the disrupted
neutron star accreting onto the black hole. The orbital and spin
energy in such a system can be tapped, for instance, through 
electromagnetic torques.

As the expanding fireball is relativistic the radiation will undergo
beaming, time-transformation, and Lorentz-boosting, blueshifting the 
emission into the gamma-ray band. 
Furthermore, the emission radiated from the blast-wave
at a given comoving time will contribute to a broad observer time 
interval, due to light travel-time effects. Fenimore et~al. (1996) 
argue that the FRED-like envelope shape of light curves are
expected from a relativistically expanding shell, and find that the decay
phase should follow a power-law.

Gamma-ray observations may give  hints of possible physical
causes of the continuum spectral emission, i.e., 
information on the processes which convert the 
kinetic energy into the observed radiation.
Empirical studies of the dynamics of the burst spectra,
enable the systematization and the investigation of the underlying
distributions of parameters. 
These empirical properties provide the important clues for the theoretical
efforts to unravel the physical processes. 
The case is probably that these relations do   not point
directly to,  and are not able to 
unambiguously state, the physical processes responsible for
the radiation. 
Several different radiation processes could be involved, as well as
pure kinematic and relativistic effects, making the physical
interpretation difficult and complex.
However, if one  of the effects is dominant, the observations
will give us direct information on physical entities, such as
the distribution of particles emitting the radiation and the optical
depth, etc. A physical GRB model must, under all circumstances, be able to 
reproduce the
severe constraints that these relations and observations give.
The large diversity, time scales, and variability of light curves must
be naturally explained, as should the shape and breaks in the
spectra. The connection between the spectral and intensity
evolution, as described above, must also be addressed by any
successful model.

The general trend of the spectrum becoming
softer over the whole burst could indicate a single emission
region having a memory of previous emission events.
 Simple radiation processes all have some difficulty in describing the
observations. 
Even on the smallest time scales the observed spectra are broad, much
broader than a black body spectrum. Is it a multi-temperature black 
body mimicking a power-law, e.g., thermal spectra from many short-lived
events  or do other processes produce the
broad spectrum? 
From the early spectral observations there were suggestions 
of thermal bremsstrahlung of an optically-thin, hot plasma.
The spectra have, however, been shown generally not to follow 
such a spectrum. The low-energy power-law does not
always follow $\alpha =-1$. Thermal bremsstrahlung is also too inefficient
(Liang 1982). Furthermore, the hardness-intensity correlation-index
is $\gamma = 0.5 -1$ for mildly relativistic thermal bremsstrahlung
from a plasma cloud (if the emission measure is constant).

A part of the internal energy will take the form of magnetic fields
which will make the relativistic electrons
radiate synchrotron radiation, a very efficient radiation mechanism.
As the fireball crashes into the
surrounding low-density gas, it will form a relativistic, collisionless
shock and radiate by optically-thin synchrotron radiation, which will
be boosted into the gamma-ray band
(the synchrotron shock model; M\'esz\'aros \& 
Rees 1993).
The electrons, giving rise to the synchrotron spectrum, are
assumed to have a truncated power-law distribution. In the comoving
frame, the minimum Lorentz factor is approximately equal to
(depending on the equipartition between electrons and protons) 
to the bulk Lorentz factor of the 
blast wave, while the maximum Lorentz factor is set by the balance of
radiative losses and energy gain from the acceleration mechanism at
work for the most energetic electrons. 
If the cooling time of the electrons is very long the
electron distribution around the low-energy cut-off will not change and
the emerging synchrotron spectrum will have a photon index of $\alpha
= -2/3$. However, if the cooling time is much shorter than the
comoving pulse
duration, the electrons will settle in a cooled distribution, emitting 
a synchrotron spectrum with a low-energy photon index between -2/3 and
-3/2, depending on the strength of the magnetic field.
The electron distribution must have a sufficiently sharp low-energy cut-off
during 
the whole evolution to be able to give rise to the observed spectrum.

The photon  index can, in this model never exceed the value $-2/3$,
creating a testable 'line of death' for the synchrotron shock model 
in its simplest version.
Preece et al. (1998b) found that 23 bursts out of their sample of 137,
for which they did time-resolved spectroscopy,
violate the synchrotron shock model, as the low-energy power-law
spectra are harder than the maximally allowed, $\alpha = -2/3$.
Cohen et~al. (1997) fitted the time-integrated spectra and found them 
to confirm the synchrotron shock model. However, as emphasized
earlier, it is the time-resolved spectra which should be studied.
The spectral index should also be constant, which clearly is not the
case, as a softening of the spectra is observed in many bursts
(Crider et~al. 1997).  
The pulse width scaling with energy, $W \sim E^{-0.45}$ is, however,
consistent with the prediction from 
radiative cooling by synchrotron losses (Tavani 1996).
Synchrotron  self-absorption would increase the low-energy power-law index,
being $+3/2$ for a non-thermal plasma.  The optical depth must then, 
however, be greater than one.

The inclusion of Comptonization of the soft synchrotron photons by the
emitting particles themselves, can modify the spectrum (Liang
et~al. 1997). The spectrum will then be an inverse Compton image of the
synchrotron continuum in the comoving frame.
The relativistic expansion will then boost the
radiation to even higher energies.
The empirical correlations point to saturated Comptonization and
as noted by Crider et~al. (1998b)
there also has to be an initial increase in Thomson depth to explain the
initial increase of $\alpha$  observed in many pulses.
Here the $W \sim E^{-0.4}$ relation, found in the $1.5 -700$ keV range, 
could be an argument against the
synchrotron self-Compton mechanism, as the X-rays and $\gamma$-rays would, 
in that case, 
be expected to have the same duration, since they would be produced by
the same population of electrons (Piro et~al. 1997).

Another emission scenario involves photon-starved Comptonization  in
an optically-thick pair plasma having moderate Lorentz factors,
$\Gamma = 30-50$. This scenario cannot 
explain the very high-energy (GeV) emission observed. It is, however,
attractive as it naturally provides a thermostat and can produce a
stable spectral break. 
The nonlinear nature 
also can explain the highly variable light curves and the large
amplitude variations, as such a system can  be turned off quickly.
See, e.g., Thompson (1998) and Ghisellini \& Celotti (1999) for more
detailed discussions.

Detailed calculations of  resulting light
curves and spectra in the blast-wave model, including synchrotron and
synchrotron self-Compton emission are given in Dermer \& Chiang (1998) and
Chiang \& Dermer (1999), showing, e.g., how the injected electron
distribution is reflected in the radiated spectrum, for various 
combinations of non-thermal electron distributions and magnetic
fields. See also, e.g., M\'esz\'aros et al. (1994) and Panaitescu
et~al. (1997).
Daigne \& Mochkovitch (1998) calculated the emitted spectrum 
in an internal shock scenario, and are able to reproduce many of the 
observations. 

An observational feature that has to be explained is the highly
variable light curves. The time scales for the variations in classical 
GRBs can be as low as $10^{-1}$ s and are found to be self-similar with
the average power density spectrum of
long bursts being a power-law over more than two 
decades of frequency (Beloborodov et~al. 1999).
Such a behavior is difficult to
explain with just variations in the external medium or shock
collisions. 
Stern (1999) suggests that complex dynamic processes in the shock
evolution make the outflow inhomogeneous, giving rise to the
observations.
This could be MHD turbulence with reconnection or
instabilities such as the Rayleigh-Taylor instability.

\section{Conclusion}

Gamma-ray bursts are observed at a rate of 1 per day with
current detectors. Their time histories are a morphological zoo
with a large diversity in shape.
The energy spectra  are peaked, broken power-laws and the 
instantaneous spectra evolve, sometimes markedly, both within 
a pulse and over the whole GRB. The intensity and its spectral
characteristics  
are often correlated. This can be described by empirical relations,
which are the result of the true intrinsic correlations of the GRB
giving rise to empirical correlations.
These are the result of the true intrinsic correlations of the GRB
event as well as of relativistic effects.
Understanding the intrinsic correlations, for instance within a burst,
will eventually lead to the unraveling of the secret behind the energy
release and the radiation processes in GRBs. 

It has been emphasized in this review  that it is important to
consider the spectral evolution when GRBs are studied, for instance, 
their light curves. Furthermore, it is the instantaneous spectra, 
which are more peaked than the time-integrated spectra, and their time
evolution that reflect the physical processes responsible for
the GRB emission. The time-integrated spectrum is,
generally, a result of the specific spectral evolution taking 
place during the burst. Unfortunately,  most theoretical spectral
models assume that it is the time-integrated spectrum that reflects 
the underlying physical emission mechanism. In the study of the 
spectral characteristics,
the low-energy power-law is the best studied and puts constraints 
on the physical radiation mechanism proposed to be responsible for the
observed radiation.
At the moment theory lags behind the observational advances. The 
observations give a number of constraints that have to be met by any
successful physical description of the GRB phenomenon.

\acknowledgments

Thanks are due to L. Borgonovo, V. Petrosian, R. Preece, J. Poutanen
and R. Svensson, for discussions and comments on the manuscript.

\end{document}